\documentclass[conference,a4paper]{IEEEtran}
\usepackage{dingbat}

\usepackage{enumitem}

\usepackage{subfigure}

\usepackage{color}

\usepackage[ruled,vlined]{algorithm2e}

\ifCLASSINFOpdf
   \usepackage[pdftex]{graphicx}
\else
\fi
\usepackage{array}
\usepackage{comment}

\usepackage{url}

\begin{document}
%
\title{Can MPTCP Secure Internet Communications from  Man-in-the-Middle Attacks?}

\author{
\IEEEauthorblockN{{Ho-Dac-Duy Nguyen}, {Chi-Dung Phung}, {Stefano Secci}, {Benevid Felix}$^*$, {Michele Nogueira}$^*$}
\IEEEauthorblockA{Sorbonne Universites, UPMC Univ Paris 06, UMR 7606, LIP6, Paris, France. Email: firstname.lastname@upmc.fr} 
\IEEEauthorblockA{$^*$Federal University of Parana, Brazil. Email: \{bfsilva,michele\}@inf.ufpr.br} 
}


%


\maketitle

\begin{abstract}
Multipath communications at the Internet scale have been a myth for a long time, with no actual protocol being deployed so that multiple paths could be taken by a same connection on the way towards an Internet destination. Recently, the Multipath Transport Control Protocol (MPTCP) extension was standardized and is undergoing a quick adoption in many use-cases, from mobile to fixed access networks, from data-centers to core networks. Among its major benefits -- i.e., reliability thanks to backup path rerouting; throughput increase thanks to link aggregation; and confidentiality thanks to harder capacity to intercept a full connection -- the latter has attracted lower attention. How interesting would it be using MPTCP to exploit multiple Internet-scale paths hence decreasing the probability of man-in-the-middle (MITM) attacks is a question to which we try to answer. By analyzing the Autonomous System (AS) level graph, we identify which countries and regions show a higher level of robustness against MITM AS-level attacks, for example due to core cable tapping or route hijacking practices.
\end{abstract}


%
\IEEEpeerreviewmaketitle

\section{Introduction}

The Multipath Transport Control Protocol (MPTCP)~\cite{RFC6824} is an extension of TCP to concurrently use multiple network paths for a given connection.
Among many proposals to support these features at the transport layer, it is considered as the one having attracted the largest interest and deployment~\cite{use-cases-RFC}. One of the main reason of this success is its incremental deployability adopted in its design, with the required signaling reusing transparently existing TCP option features. 

In MPTCP, traffic from a source to a destination in an IP network is routed over multiple \lq subflows\rq \ via different network interfaces and/or TCP ports at the transmitting and/or receiving end-points.
Subflow traffic can then be routed independently in the network segment. However, besides the usage of multiple network interfaces at the source or destination, the presence of load-balancers 
or multipath proxies~\cite{MPTCP-PM} along the network can differentiate the route followed by the subflow packets. 

Among the motivations pushed forward in support of MPTCP, there are (i) bandwidth aggregation, i.e., the increased network bandwidth offered to a connection; (ii) connection reliability, i.e., the possibility to use an alternative path in case of failure along the primary path or at the primary network interface level; (iii) communication confidentiality, i.e., the decreased ability for a Man-in-the-Middle (MITM) attacker to intercept all the traffic of a same connection.

While the first two aspects above have been largely explored in the last few years, the latter aspect is almost unexplored to date. In this paper, we report the results of an extensive measurement campaign aimed at assessing the degree of confidentiality one can expect using MPTCP. In particular, we focus on confidentiality from Autonomous System (AS)-level MITM interception, i.e., looking  at the empirical probability that a single connection can be intercepted by an organization or an attacker able to capture all the traffic going through an AS on a given direction (most of Internet communications being asymmetric).
Such attacks can happen either by remote access to routing devices of an AS or even by Border Gateway Protocol (BGP) route hijacking attacks. 
In our analysis, we consider the case of MPTCP traffic source devices using two edge providers and we compare the obtained results on a geographical basis, identifying in which countries and regions MPTCP may grant higher confidentiality with respect to MITM.
An important assumption of our analysis is that the MPTCP scheduler behavior can be modified so that it does not look for throughput maximization only, but also for path diversity exploitation for increased confidentiality, as investigated in~\cite{CNS:NOF16}.

The paper is organized as follows.
Section~\ref{background} gives a background on MPTCP and related security concerns.
In Section~\ref{metho}, we describe our measurement methodology.
Section~\ref{results} presents the results.
Section~\ref{conc} concludes the paper.

\section{Background}
\label{background}


We provide in this section the necessary background on the MultiPath TCP (MPTCP) protocol and on Internet-scale Man-In-The-Middle (MITM) attacks.

\subsection{MPTCP}

\begin{figure}[htbp]
	\centering
	\includegraphics[scale=0.2]{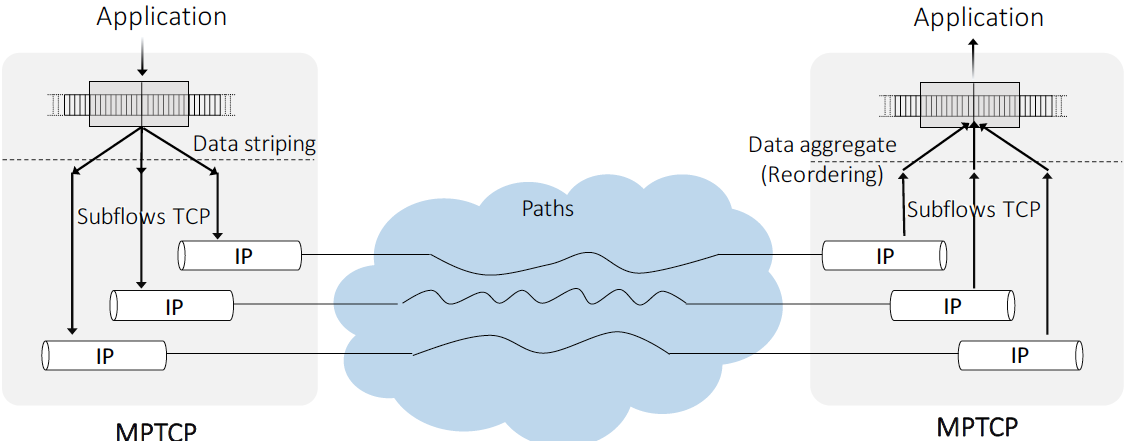}
	\vspace{-0.2cm}
	\caption{Multipath TCP Connection: Overview}
	\label{fig:mptcp}
	\vspace{-0.4cm}
\end{figure}

MPTCP extends TCP and allows to fragment a data flow from a single connection into multiple paths (subflows TCP)~\cite{RFC6824,Pen:2013}, as illustrated in Figure~\ref{fig:mptcp}. At the application layer, a connection appears as a normal TCP connection. At the network layer, each subflow looks like a regular TCP flow whose segments carry in their header a new type of TCP option~\cite{RFC6824}.  
The protocol improves the performance offered by a single flow and makes the connection more reliable using concurrent and redundant paths.

MPTCP employs a 4-tuple composed of source IP address, source port, destination IP address and destination port to identify the different subflows~\cite{RFC6824}. MPTCP manages the creation, removal and utilization of the subflows to send data while the connection is active. An MPTCP connection and its association with new subflows follow the same three-way-handshake as for initiating a normal TCP connection. In the first handshake, MPTCP uses a control flag (MP\verb+_+CAPABLE) in the option field of the segment header to verify if both end-hosts support MPTCP and configure the connection. Whether a remote host does not support the protocol, the connection seamlessy backwards to a regular TCP signaling.

The MPTCP initial handshake carries additional information, such as a cryptographic key employed to authenticate the end-hosts and set up new subflows~\cite{RFC6824}. 
In addition to the key, the establishment of additional subflow may employ other exchanged keying material, i.e., a token and random numbers (nonces), to prevent replay attacks on the authentication method. Further, an additional address identifier may be employed to identify the source IP address of a packet. Hence, even if the IP header has been changed by a middlebox (e.g. NATs, firewalls), end-hosts can unambiguously identify an address on a host.

MPTCP overcomes some weaknesses inherent to TCP, achieving (i) a \textit{greater throughput}, (ii) \textit{higher reliability}, and (iii) \textit{higher confidentiality} for data transmission than its predecessor. Indeed, a multipath connection can improve the throughput aggregating bandwidth over different paths by concurrent data transmission across all available paths. Moreover, a multipath connection can quickly overcome one path failure by sending data to another available path, increasing the data delivery reliability~\cite{Raiciu:2011}. Finally, fragmenting data flow connection across different subflows makes the flow hijacking difficult because attackers would need to capture the content transmitted through both the flows to build the content. 

Therefore, MPTCP can provide a greater confidentiality than a regular TCP transmission if the subflows of a connection are routed on disjoint paths: the higher level of disjointness, the higher the confidentiality guarantee, and the higher the robustness against such attacks. The goal of this paper is to precisely quantify this level of robustness in use-cases where MPTCP is primarily adopted not to improve communication performance or reliability, but to improve confidentiality. When addressing this feature, router-disjointness can be considered as too weak, in particular against AS-level traffic capturing and routing hijacking. We focus instead on AS-level disjointness.

\subsection{Internet MITM Attacks} 



When considering AS level path disjointness, MITM attacks can be avoided if the disjoint paths bound to physical paths  cannot jointly be accessible to a MITM attacker. In Internet-scale communications, a MITM attack can happen if the attacker is able to capture all the traffic going through a same AS. In practice, this can happen if the attacker has access to all the traffic transiting through an AS, or at least a portion of it good enough to capture traffic to or from a given prefix. This can be possible with optical layer attacks or by BGP route hijacking MITM attacks. 

At the optical layer, it is possible to split cables by using fiber optical taps, as explained in~\cite{Witcher}, with a low probability of being detected if peculiar strategies are adopted as explained in~\cite{intechwp,milcom}. Moreover, one can intercept traffic by exploiting coupling and out-of-the-fiber light propagation phenomena~\cite{coupling}, despite this is particularly challenging when performing wavelength-division-multiplexing.

At the BGP layer, a MITM attack exploits the natural way  BGP works by stealthily hijacking Internet traffic to modify  or capture it before it reaches the destination. 
 BGP MITM attacks have been quite deeply studied for about twenty years. The  survey in~\cite{Con:2016} is a recent one with a detailed description of such attacks, their effects and defenses~\cite{Con:2016}.
The BGP MITM attack type has gained special attention in 2008, when a major provider in central Asia hijacked Youtube traffic to apply local policies. In the same year, a practical BGP MITM attack was demonstrated during the DefCon hacker conference~\cite{Pil:2008}: authors  successfully intercepted traffic bound for the conference network and redirected it to a system they controlled before routing it back to DefCon. 

Multiple such BGP incidents are reported when they are notable ones~\cite{Gol:2014,Pre:2017}; often they are not reported because they cannot be always detectable, they have a limited scope, last for a short time, target a precise prefix, etc. A notable recent attack happened in 2014, when attackers have exploited such attacks to steal bitcoins~\cite{wired}: the attackers were able to inject BGP routes which redirected traffic from Bitcoin miner nodes to the attacker’s compromised host. It was estimated that at least \$83,000 worth of Bitcoins, Dogecoins, HoboNickels, and Worldcoins were stolen over a period of four months.



In~\cite{Ali:2009} a solution to protect end-to-end communications from MITM attacks is presented; it employs multiple independent paths to retrieve self-signed certificates and double-check security mechanisms during the communication. Despite it focuses on the routing layer instead of the transport layer, the work is an example of  using multiple end-to-end paths for improving security. 

At the transport layer, the advent of MPTCP has raised new specification questions and challenges. There are attempts to verify the security of MPTCP~\cite{rfc6181,rfc7430}. Also, researchers have proposed solutions against eavesdropping, in general, based on cryptography~\cite{Kim:2016}. In~\cite{rfc7430}, authors present an analysis of residual threats in the MPTCP signaling and propose fixes. In general, most of the works at the state of the art aim at either investigating security threats for MPTCP or proposing solutions for them. It is worth mentioning rising interest in using MPTCP to further enhance confidentiality when using Internet over-the-top Virtual Private Networks (VPN) services such as ToR and OnionCat~\cite{onioncat}: MPTCP is used in the upstream direction from the client to many gateways accessible via the VPN, on the way to the server, further increasing the confidentiality level of the connection. Nevertheless, such practices can have a gain which can be hardly assessed: how can you ensure the upstream source-destination traffic does follow disjoint paths, hence decrease MITM efficiency, if not at the router-level, at the AS level? In this paper, for the first time at the state of the art as of our knowledge, we attempt at providing a partial response to such questions.

\section{Methodology}
\label{metho}

In this section, we first give a description on the datasets used for constructing a representative AS-level graph of Internet, the basis for our analysis. Then, we describe our  approach for computing the number of valid vertex-disjoint paths between two arbitrary nodes over the AS graphs. Finally, we detail how we evaluate path diversity at different geographical scopes. The dataset we employed as well as our scripts are given in~\cite{lip6mptcp} for the sake of reproducibility.

\subsection{Graph construction}

The AS-level Internet connectivity graph we used is derived from two types of data: the AS-level topology data and the inter-AS relationship data, both made available in~\cite{topology}. Comparing with other resources such as~\cite{ark}~\cite{irr}, the topological data from~\cite{topology} revealed to be more reliable and able to capture a broadened view of the Internet topology.
Indeed, the data from~\cite{topology} integrates not only Routeviews  data~\cite{routeviews}, but also data from other well-known public view resources such as RIPE RIS~\cite{ripe}. Moreover,~\cite{ark} uses traceroutes that have known issues~\cite{oliveira} when converting router-level information into AS-level one. 

The inter-AS relationship data from~\cite{topology} is extracted monthly from the Cyclops database~\cite{cyclops}, which adopts the interference techniques proposed in~\cite{oliveira}. The proposed algorithm takes advantages from both BGP data and Internet eXchange Point (IXP) data, then relying on valley-free routing analysis~\cite{Gao} to derive the inter-AS relationships. To ensure the synchronization between topological and relational data we make use of measurements from the same source.

We extract 2015 data~\cite{topology} (the latest data set available), and combine them into a new dataset containing all the AS links along with their frequency of occurrence and relationship type. Employing measurements over a long period allows to capture inter-domain connection dynamics as well as inter-AS economic relationships. For instance, in one-month period, only 85\% of inter-AS links appear more than 20 days, the remaining links which have a lower frequency of occurrence being those used for backup operations or during BGP convergence periods. For the sake of consistency, we removed these unstable links. 

It is worth recalling that, because of BGP policy routing, Internet routing is asymmetric, hence capturing a full connection would mean capturing both communication directions, i.e., along both AS paths. However, as already mentioned in the introduction, we focus on MITM attacks that are able to capture traffic in one single direction, as client-server communications are essentially asymmetric.

\subsection{Path diversity computation}

The problem of selecting all the paths over a graph that satisfy given properties is often referred to as policy compliant path diversity computation in the literature ~\cite{Erlebach}~\cite{Kloti}. The general problem is to maximize the total number of valid vertex-disjoint paths in a type-of-relationship (ToR) graph~\cite{ToR}, i.e., a directed graph in which the relationship between two adjacent vertexes (ASes in our case) is expressed via the direction of the edge connecting them. A ToR graph is generated from an original undirected graph integrated with relationship data. In~\cite{Erlebach}, the path diversity between a specific pair of nodes in a graph is determined by solving an optimization formulation that maximizes the number vertex-disjoint paths in the ToR graph. The shortcoming of such approaches is that their time-complexity is relatively high hence intractable for a graph as big as the AS graph. 

The AS graph is a scale-free graph~\cite{Albert}, i.e., a graph with a relatively few hubs capturing the majority of the paths. One of the consequences is that the diameter, i.e., the length of the longest path among all the shortest paths on a direction graph, is not too high. 
However, in the AS graph the best paths are selected based on a policy routing mechanism that put before path length other criteria such as local preferences, oldest route, etc. As a result in the BGP routing table the average path length (measured in number of AS hops) is not too low neither as the best path may not be the shortest path, and it is around 5 ASes as of today~\cite{pathlength} (a bit lower in IPv6).

It is well known that searching for paths in a scale-free graph is a not too complex problem given the reasonable routing diameter, if one adopts breadth/depth-first search algorithms with a limited depth. 
Given a pair of source and destination nodes, $s$ and $d$, respectively, a graph $G(N,A)$ of policy compliant paths connecting them is discovered, where $N$ is the set of nodes and $A$ the set of links. Starting from the origin $s$, the breadth-search algorithm we adopt explores all the adjacent nodes of $s$; we validate the corresponding links $(s,n) \in A$ if they do not violate the valley-free routing property~\cite{Gao}, and label them with the corresponding inter-AS relationship.
For example, assuming that node $s$ neighbors are $n_1$, $n_2$, $n_3$, and $s$ is customer (\lq $c$\rq) of $n_1$, provider (\lq $p$\rq) of $n_2$ and peer with $n_3$; the relationships of $(s, n_1)$, $(s, n_2)$, and $(s, n_3)$ are to be labeled as \lq $c2p$\rq, \lq $p2c$\rq \ and \lq $p2p$\rq, respectively. 
Then, taking the $c$-type neighbors among $s$ neighbors (i.e., $n_2$), and looking at their neighbors $x$ in turn, those $(n_2,x)$ links are not validated if they are either $c2p$ or $p2p$ because a customer is not expected to grant transit towards its other provider(s) to one among its providers, and a customer is not expected to give access to its peer(s) to its provider(s). For instance, according to such a valley-free constraint, a valid path could be represented by the following regular expression $c2p*p2p*p2c*$~\cite{Kloti}, in which $c2p$, $p2p$ and $p2c$ express the relationship of links along a path.

Such a breath-first path search strategy is iteratively applied starting from the source, repeated for all the neighbors, neighbors of the neighbors, and so on so forth until the destination $d$ is reached or the hop distance from the $s$ gets over a given threshold $\tau$. 
By defining a proper value of $\tau$ we not only limit the time and space complexity for exploring the graph, but also ensure not taking into account long paths which should be avoid in practice by MPTCP schedulers. More importantly, the major benefit of such a search strategy is that the validity of a path can be checked in runtime while expanding the graph, thus removing edges that do not comply with routing policies.
Our search algorithm is presented in detail in Alg.~\ref{Pathsearch}.

\begin{algorithm}
\DontPrintSemicolon
\SetAlgoLined
\SetKwInOut{Input}{input}
\SetKwInOut{Output}{output}

\Input {source $s$, destination $d$, graph $g$}
\Output{ ValidPathSet }
$ VisitedNode \longleftarrow \emptyset $\;
$ queue.append([s]) $ \;
\While {queue not empty} {
    $ path \longleftarrow queue.pop() $ \;
    $ v \longleftarrow path.LastNode() $ \;
    \If {$v \neq  VisitedNode$} {
        \For {$n \in v.NeighborSet$} {
            \If {$n \neq VisitedNode$} {
                \If {label(v,n)=\lq p2c' } {
                    \For {$x \in n.NeighborSet$} {
                        \If {label(n,x)=\lq c2p' {\bf or} label(n,x)=\lq p2p'}{
                            g.RemoveEdge(n,x) 
                            }
                        }
                    }
                }
            $ NewPath \longleftarrow list(path) $ \;
            $ NewPath.append(n) $ \; 
            
            \If {n = d} {
                ValidPathSet.append(NewPath) 
            }
            \If {length(NewPath) = $\tau + 1 $  } {
                break 
            }
            queue.append(NewPath) \;
        }
        VisitedNode.add(v) \;
    }
}
\caption{Path Search Algorithm}
\label{Pathsearch}
\end{algorithm}

As a result of the path search algorithm, policy compliant paths connecting source and destination may share common nodes. To get the final set of vertex-disjoint paths, we run a simple offline filtering linear algorithm to capture the shortest disjoint paths. Since the original list of valid paths revealed to be quite small most of the times, and sorted,  the complexity of such a filtering operation is negligible.


\subsection{Source-destination pairs}
\label{sdpairs}

The current Internet ecosystem is composed of more than 50 thousands ASes, out of which more than half are stub ASes, i.e., ASes that are origin or destination only ASes. About 13\% are Tier-3 or small Tier-2 ASes we arbitrary define in this paper as those appearing at most in the third from last position and at least penultimate position in BGP AS paths; we refer to such  ASes as \lq edge provider\rq \ ASes, which  can be considered as a good representative set of national Internet Service Provider (ISPs) \lq eyeball\rq \ ASes (hence excluding Internet carriers and stub ASes).

\begin{figure}[htbp]
	\includegraphics[scale=0.4]{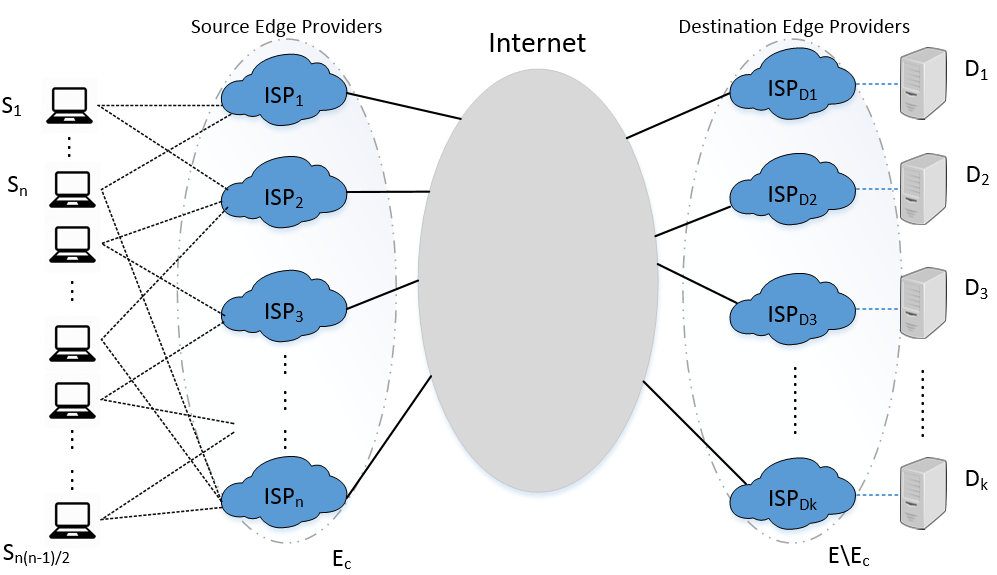}
	\caption{Source-destination pair selection process}
		\vspace{-0.3cm}
	\label{pairs}
\end{figure}

In order to determine which communications to cover in our study, we have to define a target set of source-destination pairs that address in a reasonable yet arbitrary way the communications that may be more sensible to communication privacy. 
We give a representation of the source-destination pair selection framework in Figure~\ref{pairs}.  
Our choice of source-destination pairs is as follows:
\begin{itemize}
\item the source is interconnected to two edge providers.
\item the destination is not multi-homed, i.e., it is reachable via a single ISP, the one given by the best BGP path from each source edge provider.
\item the destination belongs to another AS and another country than the ones of the source.
\end{itemize}

Besides reducing the number of pairs to a reasonable and treatable number (requiring about one week of computation), it is worth noting that, in such a way, we consider one single-direction communication data: the data we care of is the one is sent from the source node to the destination node. That is, under such a path election strategy, we are covering:
\begin{enumerate}
\item the case of multi-homed (nomadic) devices \textit{uploading} to single-homed server;
\item the case of single-homed (fixed) devices \textit{downloading} contents from servers connected to a multi-homed network.
 \end{enumerate}

For instance, the first case may be the one of sensors or other mobile devices (using cellular and/or WiFi and/or IoT protocol interfaces) uploading to a private cloud storage collected data (measurements, photos, videos, etc), and the second case may be the one of a residential user downloading files at home.
Therefore, we are not covering:
\begin{itemize}
\item the case of multi-homed (nomadic) devices \textit{downloading} from single-homed server;
\item the case of single-homed (fixed) devices \textit{uploading} contents to servers connected to a multi-homed network.
\item the case of multi-homed devices communicating with another multi-homed device;
\end{itemize}

We do not want to speculate that the latter three cases are less likely to happen. A dual analysis, quite  expensive computationally, covering these cases may be performed as well in future works. However, at this stage, we cannot see intuitive critical-communication use-cases corresponding to such cases, and we doubt such an extended analysis might bring additional significant insights.

Let us more precisely describe the source-destination pair election process. 
We segment the set of edge providers, $E$, in country-specific subsets, $E_c$, where $c \in C$ is a country in the set of countries $C$, i.e., $E = \cup_{c\in C} E_c$.
We use the AS-to-country mapping given by the CIDR Report~\cite{country}.
Overall, for a given country $\tilde{c}$, the number of source-destination pairs is therefore equal to:
\begin{equation}
     \frac{ \vert E_{\tilde{c}} \vert \times (\vert E_{\tilde{c}} \vert - 1) }{2} \times \sum_{c \neq \tilde{c}} \vert E_c\vert 
     \label{sdpair}
\end{equation}

Doing so, we are targeting a lower bound, pessimistic, analysis, because we consider only international communications and we suppose the destination is not multi-homed. 
The filter we set on the destination enumeration allows us to target communications that may need a higher confidentiality due to their international connotation. Moreover, in this way we also avoid a huge bias due to the fact that a large majority of the AS paths available at the national level are not visible in backbone BGP routing tables such as the Routeviews ones (typically because of internet exchange points, as recently shown in~\cite{ixproutes}). 
We believe having a lower bound stand is more appropriate than an upper bound one, while allowing us to scientifically qualify the value of the relative trends.

Given a source-destination pair, we  then calculate the number of AS-disjoint paths using the path search algorithm already described, and store this information. In practice, on the resulting AS-graph (composed of 21469 edge provider nodes and 86983 edges with an average degree of 4.05), it takes from 500~ms to 2.5~s to compute the total number of policy-compliant vertex-disjoint paths for each pair of AS nodes.
Please note that the source node is not considered as an AS node of the AS graph; therefore, in the measurement results that follow we distinguish between the view point of the source node from the view point of its providers.

\section{Results}
\label{results}

\begin{figure}[!hb]
	\centering
		\subfigure[device view]{\includegraphics[width=8.8cm]{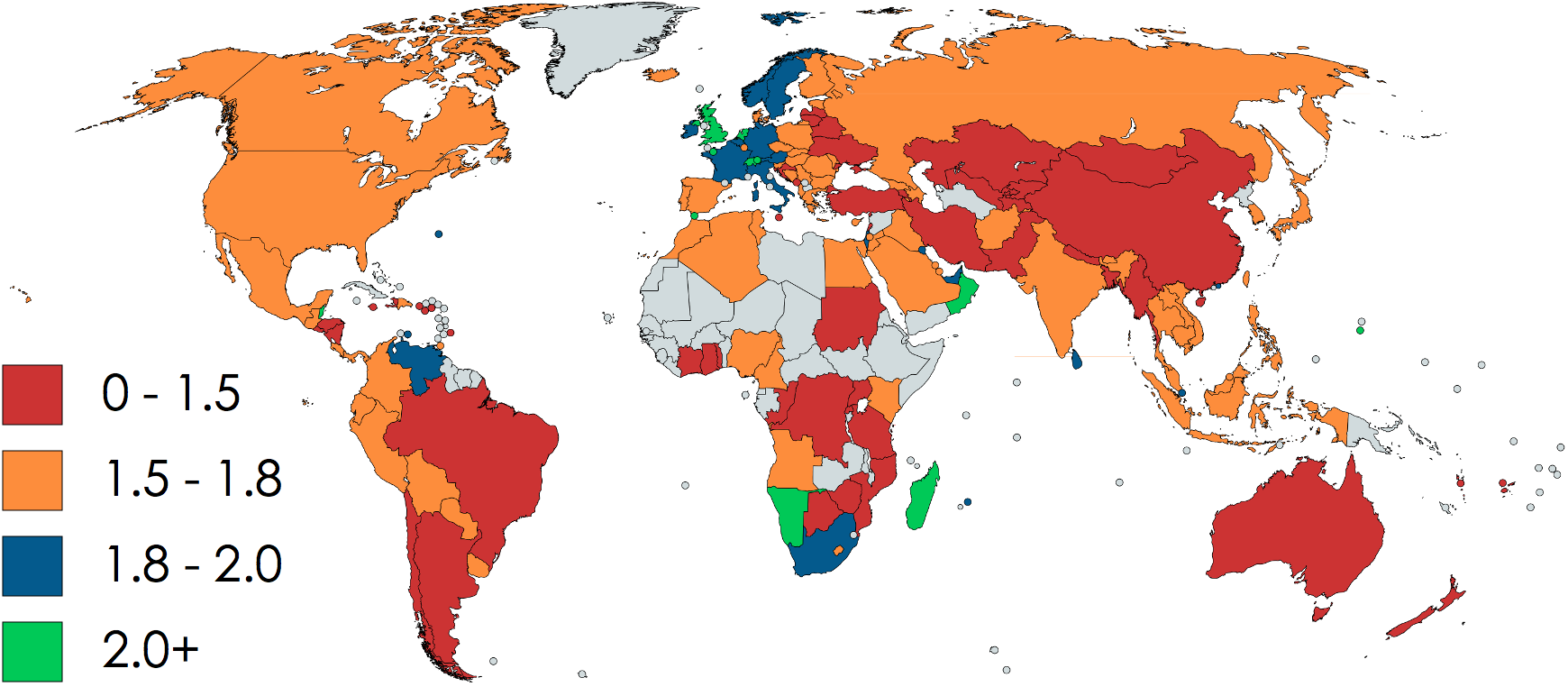}}
	\subfigure[edge provider view]{\includegraphics[width=8.8cm]{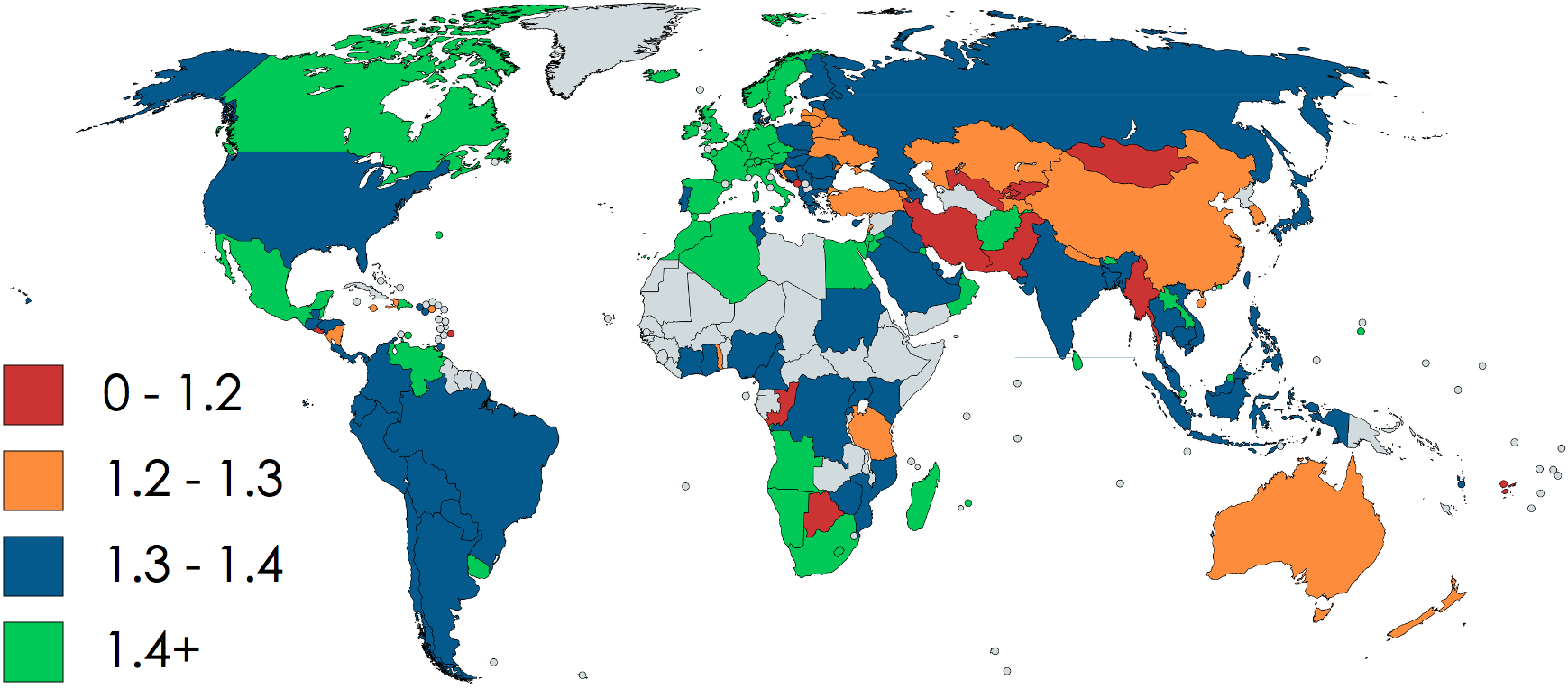}}
	\caption{Countries covered with corresponding MITM robustness distribution.}
	\label{countriesmap}
\end{figure}

We report the results obtained for a set of 147 countries, i.e., those countries from the United Nations statistics~\cite{un} that have at least two distinct edge providers officially based in the country (this excludes Greenland territories, very small city-state countries and many African and Indonesian countries). The geographical coverage is given in Figure~\ref{countriesmap}.

\begin{figure*}[htbp]
	\includegraphics[scale=0.36]{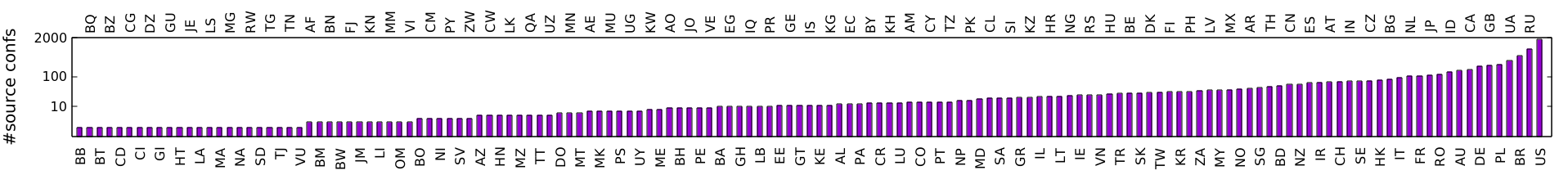}
	\vspace{-1cm}
	\caption{Number of source configurations per country}
	\vspace{-0.4cm}
	\label{cdfpairs}
\end{figure*}

\begin{figure*}[htbp]
	\includegraphics[scale=0.36]{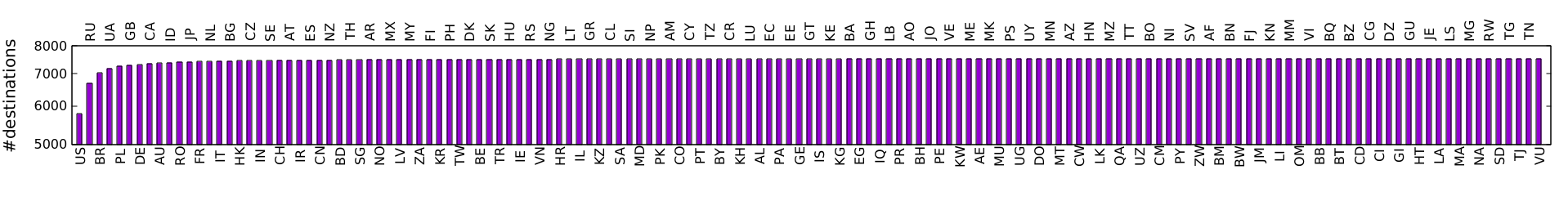}
	\vspace{-1cm}
	\caption{Size of  the destination set per country}
	\vspace{-0.5cm}
	\label{cdfdest}
\end{figure*}

We proceed as follows.
\begin{itemize}
    \item For each country, we generate all possible dual-homed source configurations, i.e., all possible pairs of edge providers. Figure~\ref{cdfpairs} reports the distribution of the number of source pairs for each countries.
    \item For each source configuration (i.e., for each pair of edge providers), we compute the number of disjoint paths to each destination. Figure~\ref{cdfdest} reports the distribution of the number of destinations for each source country.
    \item \textit{MITM robustness} metric: for a given source, we define it as the average of the number of disjoint paths over all the destinations: such metric can be considered as a level of unlikelihood that a MinM attack can take place for that source configuration; the higher the value of the MITM metric, the more difficult is for an attacker to capture traffic from that country.
    \item For each country, a series of MITM robustness metrics, one for each source, is therefore created.
\end{itemize}

We characterize the resulting series using  boxplot (minimum, quantiles, maximum, and 0.1\% outliers) distribution.
Figures~\ref{cdfpairs}~and~\ref{cdfdest} show that the three countries with the highest number of source configurations (i.e., those with the highest number of edge provider pairs), e.g., Brazil, US and Russia have also the lowest size of the destination set, which is reasonable given (\ref{sdpair}), while guaranteeing a largely sufficient statistical significance (thousands of entries for each country).
We overlay over the boxplots the average of the corresponding series (red square). The results are then ordered from left to right with increasing averages.

\begin{figure*}[htbp]
		\subfigure[device view]{\includegraphics[width=18.2cm]{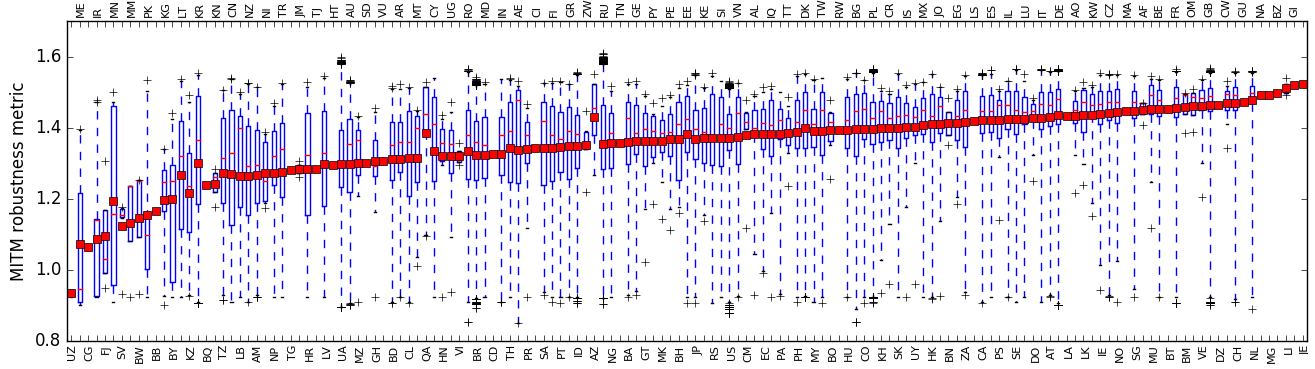}}
	\subfigure[edge provider view]{\includegraphics[width=18.2cm]{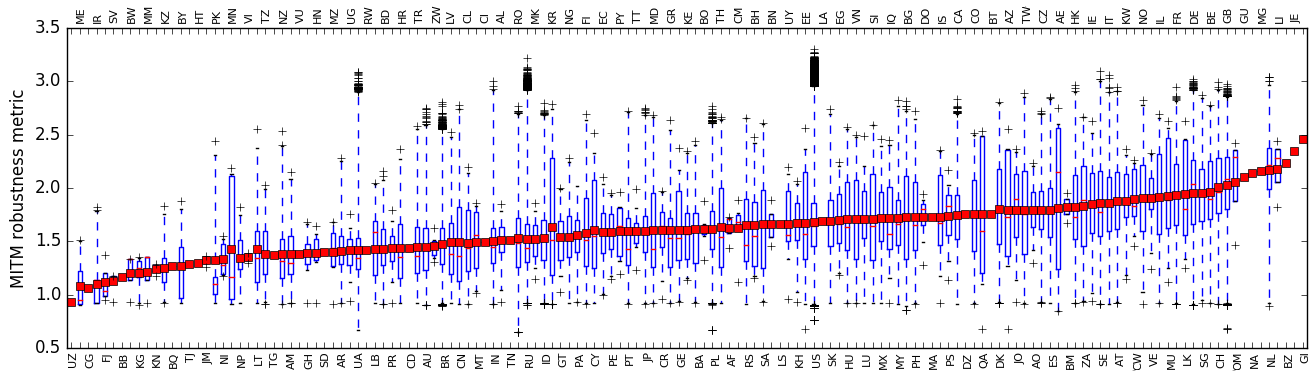}}
	\subfigure[differential robustness view]{\includegraphics[width=18.2cm]{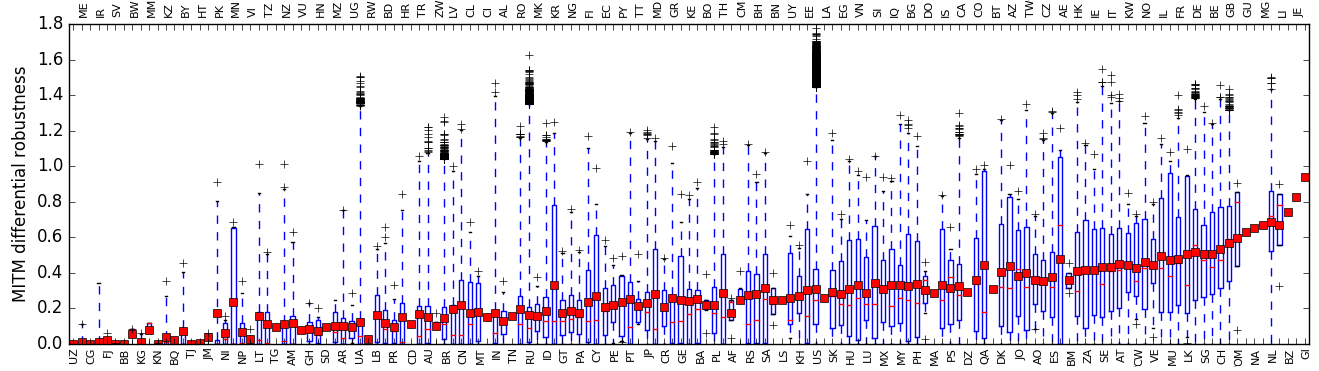}}
	\caption{MITM robustness distribution for 147 countries.}
	\vspace{-0.6cm}
	\label{countries}
\end{figure*}

Results are shown in Figure~\ref{countries} and depicted on the world map in Figure~\ref{countriesmap}.
We report three types of results:
\begin{itemize}

\item
For Figure~\ref{countries}a, the MITM robustness is computed with the source node integrated as a node in the AS graph as an artificial node, i.e., it provides a \textit{device view}; obviously, in this view the upper bound of the robustness is 2, i.e., the number of edge providers used by the source. 
\item 
For Figure~\ref{countries}b, the MITM robustness is computed instead summing the number of disjoint paths from the first edge provider to those from the second edge provider, then decreased by those paths that do share an AS hop, i.e., it provides an \textit{edge provider view}; when higher than 1 for a given edge provider, our assumption when counting the robustness metric is that the additional paths from the edge provider can be made available to MPTCP subflows by forms of  load-balancing.
\item For Figure~\ref{countries}c, we report the differential robustness results, i.e., the edge provider view robustness minus the device view robustness, computed for each source configuration individually. This view more precisely quantifies the gain achievable for MPTCP communications in case of inter-AS load-balancing at the edge providers.
\end{itemize}

That is, while Figure~\ref{countries}b assumes MITM attacks do not happen at the source and destination edge providers (i.e., there is a high level of trust on those edge providers), Figure~\ref{countries}a assumes that MITM attacks can happen at the source edge providers, hence revealing a low level of trust in source direct providers. 
Figure~\ref{countries}c can be interpreted as the marginal gain to consider when adopting (for the provider) or using (for the user) inter-AS load-balancing features at the edge provider (or over-the-top VPN) level.
It is worth noting that Figure~\ref{countries}a may also better represent the case of single-homed client downloading from multi-homed server discussed in Section~\ref{sdpairs}, in which case server direct providers are not the client direct providers and thus can be less trusted ones.

As a general assessment, Figures~\ref{countries} show that, considering 1.5 average as the rough threshold making the likelihood of MITM negative if higher than it, positive is lower than it,  only about 5\% of the countries show good chances of being robust against MITM from a device view, while looking at the maximum instead of the average and median values one could speculate that careful choice of the edge providers could make this likelihood positive for a majority of the countries. From a, edge provider view this ratio grows to roughly 60\%, and higher than 90\% if the edge provider choice is influenced by confidentiality concerns.

Moreover, the average number of paths connecting a dual-home node to international destinations has a significant variance depending on the source country. The average robustness ranges from 1 and less to 1.6 from a device view, and from 1 (and less) to 2.5 from an edge provider view. Some minimum and even average values are below 1 because of the partial view of Internet topology and the incompleteness of inter-AS relationship interference, which make some destinations unreachable via the ToR graph (hence counted as 0 paths); we left it as is to also give an index of the level of the ToR graph incompleteness for different countries. In any case, the boxplot median is a metric robust against such outliers to look at.

In addition, observing the plots in Figure~\ref{countries}, we can also remark that:
\begin{itemize}
\item 
Within a country, a high inter-quantile range (between the first and the third quantiles) indicates that the path diversity strongly depends on how the two upstream edge providers are selected for the source.

\item
The median is mostly higher than the average in the device view, and lower than the average in the edge provider view. This is essentially due to outliers, counted in the average and not in the median.

\item The gap between the min and max robustness is another interesting fitness metric to observe. Some countries maintain a small gap (below 1) while others have a very big gap (up to 2). In other words, the deployment of MPTCP for securing international communications in some countries can statistically yield a much better result than in other countries, where this gap is smaller. Particularly interesting is the case of Angola (AO), Venezuela (VE) and Namibia (NA), with small robustness gap, which may be correlated to the presence of inter-continental cables landing at or close to the country~\cite{cableinfo}.  

\item In the edge provider view, the maximum value is higher than 2 in the majority of the countries: this behavior suggests that with a proper choice of trusted upstream providers for the source, communicating nodes can adopt MPTCP to statistically expect high  confidentiality for their communications regardless. Particularly alerting are the cases of Uzbekistan (UZ), Nepal (NP) and Lebanon (LB), with quite low maximum value.

\item In the device view, in the majority of the cases the maximum robustness is not higher than 1.6, both averages and medians are quite far from the desirable target of 2. Hence, without the support of inter-AS load-balancing (such as with BGP or LISP protocols), path diversity from a dual-homed node is reduced significantly, indicating a non negligible probability of paths joining on the way to the destination.  


\item Looking at the differential robustness, we can remark that among the countries that have the lowest device view MITM robustness, those that could the most benefit from inter-AS load-balancing practices are Mongolia (MN), Pakistan (PK) and Korea (KR). However, most of those with low robustness do not improve much the situation with respect to the device view.
\end{itemize}

\begin{figure}[htbp]
	\subfigure[\underline{device view}:  macro-regions grouping]{\includegraphics[width=6cm]{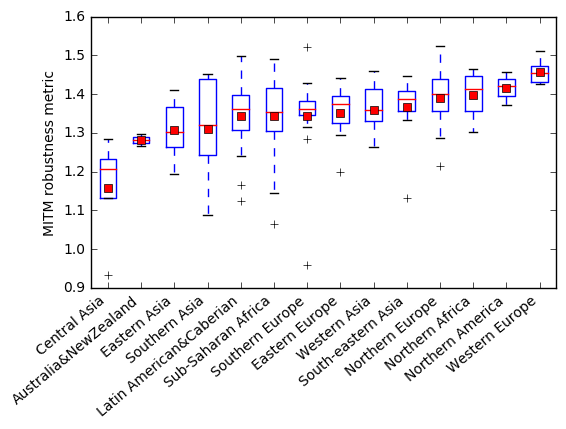}}
	\subfigure[position grouping]{\includegraphics[width=2.65cm]{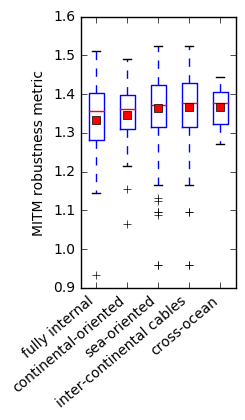}} 
	\subfigure[\underline{edge provider view}:  macro-regions grouping]{\includegraphics[width=6cm]{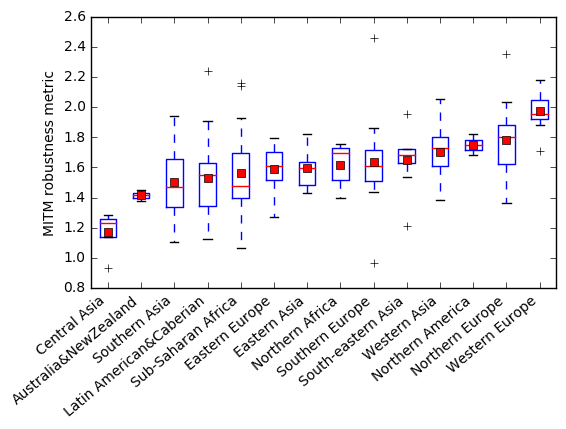}}	
	\subfigure[position grouping]{\includegraphics[width=2.65cm]{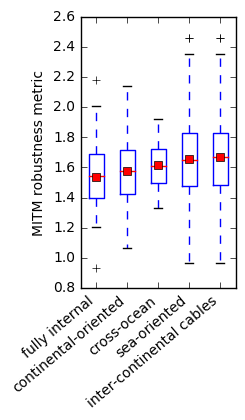}}
	\caption{MITM robustness metric with aggregated country groups.} 
	\label{regions}
\end{figure}

Looking at macro geographical regions, many European countries seem to grant better security than countries in other regions. 
In order to look after continental characteristics, the plots in Figure~\ref{regions} show the boxplot results (with 1\% outliers) aggregated on a macro-region basis (a and c, sub-continental level) and on a relative position basis (b and d, in terms of seacoast and inland borders).  
We can remark that:  
\begin{itemize}
 
\item Western Europe appears as the best off, followed by Northern Europe and Northern America. In almost 50\% of Western Europe countries there can be 2 disjoint paths from the source edge providers to Internet destinations.
\item 
Central Asia shows the worst robustness, followed by Australia and New Zealand; the reasons are likely network centralization practices  and geographical isolation. It is interesting to notice the relevant gap between Central and South-Eastern/Western Asia.
\item A high variance is recorded at Southern Asia, Northern Europe and Sub-Saharan Africa, which indicates high  differences among the countries within these areas. 

\item We could not find strong correlation between the relative continental position, and the robustness metric, yet a positive correlation exists, with countries at the boundaries of oceans, with inter-continental cable landing and that are sea-oriented (most of the border on the sea-side) that offer higher robustness than fully internal and continental-oriented ones.

\end{itemize}

\section{Application scopes}

We focused our study on MPTCP-based communications. More precisely, our study cover the following cases:

\begin{itemize}
\item \textit{MPTCP capable end-points}: both source and destination, client and server (or viceversa) end-points, are MPTCP capable, and the MPTCP communication is not filtered by middle-boxes. As argued in Section~\ref{sdpairs}, the multi-homed end-point can be either the server or the client.
\item \textit{MPTCP proxied end-points}: at least one end-point is not MPTCP capable, but its (their) TCP communications is (are) handled by MPTCP proxie(s), converting TCP packets into MPTCP packets and/or viceversa, as explained in~\cite{MPTCP-PM,MPTCP-NAM}, possibly routed via Internet disjoint paths as proposed in~\cite{CSPRG:CLOUDNET13,BPS:ANCS15}. The multipath conversion proxies (i.e., aggregation and concentrator nodes) can sit at end-point premises (customer premises equipment for the client node, hypervisor or middle-box at the server side) or at the edge provider level borders.
\end{itemize}

Moreover, besides MPTCP-based communications, other protocols offering Internet-scale multipath load-balancing could also be covered by this study. The following protocols are either not deployed, or they have undergone only a limited deployment so far; they are:

\begin{itemize}

\item \textit{SCTP}: the Stream-Control-Protocol (SCTP)~\cite{sctprfc} is another multipath transport protocol absolving the same function than MPTCP, but less deployed than MPTCP due to the limited retrocompatibility.

\item \textit{LISP}: the Locator/Identifier Separation Protocol (LISP)~\cite{lisprfc} is able to perform inter-AS inbound load-balancing by means of encapsulation, routing locator mapping, and appropriate traffic engineering policy configuration.  LISP primary scope is the edge provider one, hence results with the edge provider view readily apply to LISP traffic engineering. It is worth noting that, deployment of LISP as an intra-AS TE tool can also allow performing inter-AS multipath on the outbound direction as proposed in~\cite{xavierLISPTE}.

\item \textit{MultiPath BGP}: in BGP, when some higher BGP decision criteria are equivalent, even if the load might technically be balanced on the equivalent routes, only one route is retained using lower-level criteria (that can be inefficient ones in terms of global routing such as hot-potato or tie-breaking rules). However, forms of \textit{Multipath BGP} were discussed in standardization fora, but finally not standardized; however, some recommendations have been published~\cite{bgp4issues}, and some vendors implement it (see, e.g.,~\cite{junipermp} and~\cite{ciscomp}). 
Such multipath BGP mode can be adopted at the edge provider scope as for LISP, but instead than on the inbound direction as LISP, multipath BGP applies to the egress load-balancing direction.
Despite the study~\cite{ERS:NGI-10} on BGP core routing tables report that in 2010 multipath BGP was practically not used, speculations report it is used by major cloud providers.

\end{itemize}

The above protocols are a selection of those protocol communication contexts where load-balancing can have a direct effect on the AS path selection.
Nevertheless, other load-balancing protocols  can potentially have an impact on the egress AS selection as well, as for instance in data-center environments. 
In the case of MPTCP communications, these protocols, operated at the edge provider view, are  able to perform inter-AS load-balancing in such a way that the path diversity exposed in the edge provider view in our analysis can be made available to MPTCP devices subflows, hence given them the full potential of MPTCP in terms of communication confidentiality and robustness against MITM attacks.

\section{Conclusion}
\label{conc}

We explored in this paper how Internet path diversity could be exploited by means of multi-path transport protocols when looking at increased security against man-in-the-middle attacks. We focused in particular on such attacks acting at the Internet network autonomous system level, and at the robustness of MPTCP communications in what appear as a reasonable configuration where at least one among the endpoints is multi-homed with two edge providers.

We reported extensive specific and aggregated results for most of the world countries and regions, looking at macro trends that could inspire further research in the area. Results show that, statistically speaking, MPTCP does not help in guaranteeing robustness against MITM attacks hence high confidentiality, unless the choice of the edge provider is carefully taken, or one can rely on inter-AS load-balancing features offered implicitly or explicitly by edge providers. Some continental regions are strongly more robust than others, and there seems to be a positive correlation with inter-continental cable landing proximity. Moreover, the results show  that there are countries surprisingly less well connected than one could think of and countries that are more obviously less robust against such attacks due to network centralization practices.




\end{document}